\documentclass[smallextended]{svjour3}     

\smartqed  
\usepackage{graphicx}
\usepackage{amsmath}     
\usepackage{subfig}      

\renewcommand{\tilde}{\widetilde}
\renewcommand{\hat}{\widehat}
\renewcommand{\d}{\mbox{d}} 
\renewcommand{\i}{\mbox{i}} 
\newcommand{\e}{\mbox{e}}  
\newcommand{\rac}{\sqrt{(y^2+1)^2+\mu^{-2}}}

\newcommand{\hI}{\text{\it \^I}}   

\newcommand{\Orz}{\Omega_0} 
\newcommand{\am}{\mbox{am}}
\newcommand{\sn}{\mbox{sn}}
\newcommand{\cn}{\mbox{cn}}
\newcommand{\dn}{\mbox{dn}}
\newcommand{\K}{\mbox{K}}
\newcommand{\F}{\mbox{F}}
\newcommand{\E}{\mbox{E}}

\newcommand{\mtz}{\tilde{M}}     
\newcommand{\dOz}{\tilde{\zeta}} 
\newcommand{\zsM}{\hat{\zeta}}   
\newcommand{\vpd}{\upsilon}
\newcommand{\muK}{(\mu-\mu_0)}   
\newcommand{\nn}{\nonumber \\}

\journalname{General Relativity and Gravitation}

\begin{document}

\title{On the black hole limit of rotating discs and rings}

\dedication{Dedicated to Gernot Neugebauer on the occasion of his 70th birthday}

\author{ A.~Kleinw\"achter \and H.~Labranche \and R.~Meinel}

\authorrunning{A.~Kleinw\"achter et al.} 

\institute{
A.\ Kleinw\"achter \and
H.\ Labranche \and
R.\ Meinel 
\at Theoretisch-Physikalisches Institut, University of Jena,\\
Max-Wien-Platz 1, 07743 Jena, Germany\\
\email{meinel@tpi.uni-jena.de}
}

\date{Received: date / Accepted: date}

\maketitle

\begin{abstract}
Solutions to Einstein's field equations describing rotating fluid bodies in equilibrium permit parametric (i.e.\ quasi-stationary) transitions to the extreme Kerr solution (outside the horizon). This has been shown analytically for discs of dust and numerically for ring solutions with various equations of state. From the exterior point of view, this transition can be interpreted as a (quasi) black hole limit. All gravitational multipole moments assume precisely the values of an extremal Kerr black hole in the limit. In the present paper, the way in which the black hole limit is approached is investigated in more detail by means of a parametric Taylor series expansion of the exact solution describing a rigidly rotating disc of dust. Combined with numerical calculations for ring solutions our results indicate an interesting universal behaviour of the multipole moments near the black hole limit.  
\end{abstract}

\section{Introduction}
A fascinating property of some equilibrium configurations of rotating fluid bodies is the existence of a ``black hole limit'' when a well-defined parameter relation is reached. This was first demonstrated by Bardeen and Wagoner \cite{bw} with their approximate solution to the problem of a uniformly rotating disc of dust in general relativity. The exact solution to the disc problem found by Neugebauer and Meinel \cite{NM93,NM95} confirmed this result, see \cite{m02,MAKNP08} for details. In the limit, the spacetime separates: From the ``exterior point of view'', the extreme Kerr metric outside the event horizon is formed. From the ``interior point of view'', a regular, non-asymptotically flat spacetime -- containing the fluid body -- with the extreme Kerr ``near-horizon geometry'' (also called ``throat geometry'' \cite{bh}) at spatial infinity results. Similar phenomena were observed for limiting solutions to the static Einstein-Yang-Mills-Higgs and Einstein-Maxwell equations \cite{bfm,lw,lz07,lz10,bonnor}. Strictly speaking, there is not yet a horizon in the limit, and hence the denotation ``quasi-black hole'' used by Lemos {\it et al.} \cite{lz07} is also appropriate. However, as already discussed by Bardeen \cite{Bardeen}, the slightest dynamical perturbation will lead to a genuine black hole. Therefore, we continue to use the term ``black hole limit''. Such a limit was also found numerically for uniformly rotating fluid rings with various equations of state \cite{akm,fha05,LPA07}. 

It was shown in \cite{m04} that the black hole limit of a (uniformly) rotating fluid body in equilibrium always leads to an extremal Kerr black hole\footnote{See \cite{m09} for a discussion of the extension of the Kerr uniqueness proof to the case with a degenerate horizon.}. A necessary and sufficient condition for the limit is the parameter relation \cite{m06}
\begin{equation}\label{BHcond}
M - 2\Omega J \to 0, 
\end{equation}
where $M$ denotes the total (gravitational) mass, $J$ the angular momentum and $\Omega$ the angular velocity\footnote{Throughout the paper we use units in which the speed of light as well as Newton's gravitational constant are equal to $1$.}. Note that Lemos and Zaslavskii \cite{lz09} have shown in a more general context that quasi-black holes are always extremal if infinite surface stresses are excluded.   

In \cite{LPA07} evidence for a universal behaviour of the Geroch-Hansen gravitational multipole moments \cite{Geroch70,Hansen74} near the black hole limit of rotating fluid bodies was reported. In the present paper, these results are confirmed and extended. In particular, for the disc of dust, a parametric Taylor series expansion of the exact solution at the black hole limit is derived. The multipole moments can be calculated according to the algorithm introduced by Fodor {\it et al.} \cite{FHP89} using the Ernst formulation of the stationary and axisymmetric vacuum Einstein equations \cite{Ernst68,KraNeu68}.    

\section{The Ernst potential of the rigidly rotating disc of dust}
\subsection{The metric and the Ernst potential}
The asymptotically flat spacetime of a rigidly rotating disc of dust surrounded by a vacuum can be described using the metric
\begin{equation}
\d s^2=\e^{-2U}[\e^{2k}(\d\rho^2+\d\zeta^2) +\rho^2 \d\varphi^2]-\e^{2U}(a \d\varphi +\d t)^2 ~, \label{metric-Weyl}
\end{equation}
where the functions $\e^{2k}$, $\e^{2U}$ and $a$ depend only on the canonical Weyl coordinates $\rho$ and $\zeta$, and not on the time coordinate $t$ and the azimuthal angle $\varphi$, corresponding to stationarity and axial symmetry.
The equatorial ``plane'' is given by $\zeta=0$ and the axis of rotation by $\rho=0$.

Einstein's vacuum field equations, in the case of stationarity and axial symmetry, are equivalent to the complex Ernst equation \cite{Ernst68,KraNeu68}
\begin{equation}
(\Re f) \nabla^2 f = \nabla f \cdot \nabla f~, \label{ernst-eqn}
\end{equation}
where $f(\rho,\zeta)$ is a complex function called the Ernst potential,
and $\nabla^2$ and $\nabla$ are respectively the Laplace and the gradient
operators in a three dimensional Euclidean space, as if $\rho$, $\zeta$ and $\varphi$
were cylindrical coordinates.
The metric potentials can then be calculated from:
\begin{eqnarray}
e^{2U}     & = &  \Re f ~,\\
a,_{\rho}  & = &  \rho \e^{-4U} b,_{\zeta} ~,\\
a,_{\zeta} & = & -\rho \e^{-4U} b,_{\rho} ~,\\
k,_{\rho}  & = &  \rho [U,_{\rho}^2-U,_{\zeta}^2+ \frac{\e^{-4U}}{4} (b,_{\rho}^2-b,_{\zeta}^2)] ~,\\
k,_{\zeta} & = & 2\rho [U,_{\rho} U,_{\zeta}+\frac{\e^{-4U}}{4} b,_{\rho} b,_{\zeta}] ~,
\end{eqnarray}
where $b(\rho,\zeta)$ is the imaginary part of the Ernst potential $b:=\Im f$ (so $f=\e^{2U}+\i\, b$).
For asymptotically flat solutions with the gravitational mass $M$ and angular momentum $J$,
the behaviour of the Ernst potential at infinity is
\begin{equation}
f=1-\frac{2M}{r}+\frac{2(M^2-\i J\cos \theta)}{r^2} + {\cal O}\big(\frac{1}{r^3}\big)~, \label{f-at-infinity}
\end{equation}
where $r:=\sqrt{\rho^2+\zeta^2}$ and $\tan\theta:=\rho/\zeta$.

\subsection{The rigidly rotating disc of dust}
\begin{figure}
\begin{center}
 \includegraphics{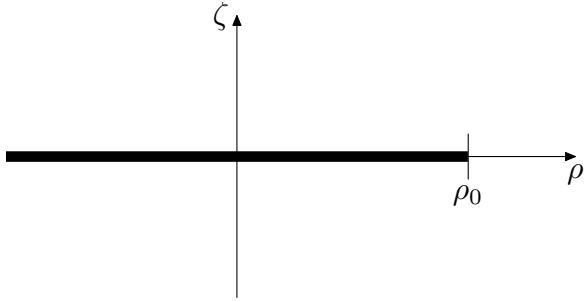}
 \caption{The thick line is the infinitesimally thin disc of dust. The disc has a radius $\rho_0$ and rotates about the $\zeta$-axis.} \label{fig-disk_of_dust}
\end{center}
\end{figure}
The thin disc of dust is represented in Fig.~\ref{fig-disk_of_dust}. It is located at $\zeta=0$, $\rho\le\rho_0$ and rotates with the constant angular velocity 
$\Omega$ with respect to infinity about the $\zeta$-axis. It is useful to introduce a potential $V$ by
\begin{equation*}
  \e^{2V} = \e^{2U} [(1+\Omega a)^2 - \Omega^2 \rho^2 \e^{-4U}]~.
\end{equation*}
For any uniformly rotating fluid body in equilibrium\footnote{While for the disc of dust the metric \eqref{metric-Weyl} can be used globally, it can only be introduced in the exterior (vacuum) region in case of genuine fluid bodies.} the function $V$ is constant ($V\equiv V_0$)
along the surface of the body (the disc ``consists'' of a surface only). This constant $V_0$ is related to the relative redshift $Z_0$ of photons (without $\zeta$-component of orbital angular momentum) emitted from the surface and received at infinity via
\begin{equation}\label{redshift}
 \e^{-V_0} -1 = Z_0~.
\end{equation}
The surface mass density is not constant within the disc:
it must vary in a unique way such 
that each particle of dust can follow a geodesic of constant $\rho$, all with the same $\Omega$, by the sole effect of gravitation. The physical parameters that we just mentioned 
can be combined into a ``relativistic parameter'' called $\mu$:
\begin{equation}
\mu:=2\,\Omega^2\rho_0^2\,\e^{-2V_0}~.
\end{equation}
The disc has a set of solutions that can be specified by fixing two parameters. But
by using normalized and dimensionless coordinates, the disc can be characterized by a single parameter.
Thus, from its Ernst potential we can extract dimensionless combinations of 
physical parameters as functions of $\mu$ alone. The gravitational mass $M$, the angular momentum $J$,
the imaginary part $b_0$ of the Ernst potential at the centre of the disc ($\rho=0$, $\zeta=0^+$) and the previously introduced parameters
are given by
\begin{subequations} \label{functions-of-mu}
\begin{align}
\e^{2V_0} &= \frac{h'\cn^2(\hI,h')}{h}~, \\
 b_0 &= -\frac{\sn(\hI,h')\; \dn(\hI,h')}{h} ~,\\
\Orz \equiv \Omega \rho_0 &= \frac{1}{2}\sqrt{1-\frac{h'^2}{h^2}}\;\cn(\hI,h') ~,\label{func-Om0}\\
\mtz     \equiv 2\Omega   M &= -b_0- \Orz c_1~, \\
\tilde{J}\equiv 4\Omega^2 J &= -b_0-2\Orz c_1~,
\end{align}
where $\sn(u,k)$, $\cn(u,k)$ and $\dn(u,k)$ are the Jacobian elliptic functions, 
and $h$, $h'$, $\hI$ and $c_1$ are the following functions of $\mu$:
\begin{align}
h =~& \sqrt{\frac{1}{2}\Bigl( 1+\frac{1}{\sqrt{\mu^{-2}+1}}\Bigl)}~, \quad\quad
h'  = \sqrt{\frac{1}{2}\Bigl( 1-\frac{1}{\sqrt{\mu^{-2}+1}}\Bigl)}~,\\
\hI =~& \frac{\sqrt[4]{1+\mu^2}}{\pi}\int^{\mu}_0 \frac{g(x)}{\sqrt{\mu-x}}\d x ~,\quad\quad g(x)=\frac{\ln(\sqrt{1+x^2}+x)}{\sqrt{1+x^2}}~,\\
c_1 =~& \frac{1}{\sqrt{\mu}} \Bigl\{ 2\sqrt[4]{1+\mu^2}\; \E(\am(\hI,h'),h') - (\mu+\sqrt{1+\mu^2})I_0 +I_1\Bigr\} \label{c1}\\
\text{and}\quad I_n =~& \frac{1}{\pi}\int^{\mu}_0 \frac{g(x)x^n}{\sqrt{\mu-x}}\d x~. 
\end{align}
\end{subequations}
The Ernst potential for a uniformly rotating disc of dust is given in \cite{NKM96,NM03} for the whole spacetime in terms of hyperelliptic theta functions and quadratures.
Since the multipole moments of stationary and axisymmetric configurations -- which contain all the information about the exterior gravitational field -- can be obtained from the Ernst potential on the symmetry axis, it is sufficient to restrict ourselves to this axis, where the potential becomes simpler, see \cite{NM94}. The representation 
of the potential that we will use is considerably different from the one given in \cite{NM94},
but all the necessary relations are provided in detail in \cite{K00,MAKNP08}.

To make the Ernst potential of the disc a function of one physical parameter, in our case $\mu$, 
we begin by introducing the normalized and dimensionless coordinates $x:=\rho/\rho_0$ and $y:=\zeta/\rho_0$.
The potential on the axis is obtained by evaluating it in the limit $x\to 0$;
it then becomes a function of two variables: $f(\mu;x=0,y)\to f(\mu;y)$. 
By restricting ourselves to the positive part of the axis ($y>0$),
the Ernst potential can be written as follows:
\begin{subequations}\label{functions-of-mu-and-y}
\begin{equation}\label{f_of_disk}
f(\mu;y) = \frac{1-\i NQ_-}{N+\i Q_+} 
\end{equation}
with the following functions defined as real functions of $\mu$ and $y$:
\begin{align}
Q_\pm =~& \frac{1-2\Orz^2(y^2+1+\rac)}{-b_0 \mp 2\Orz y} ~,\\
N =~& \exp \Bigl\{ L-2\hI(S+H+Z+A)+T \Bigr\} ~,\\
L =~& \frac{y\sqrt{\mu}}{\pi\rac} \int_0^\mu \frac{x(y^2+1)+\mu^{-1}}{\mu(y^2+1)-x} \cdot\frac{g(x)}{\sqrt{\mu-x}}\d x ~,\\
S =~& \frac{h'(y^2-\tau^2)}{2\rac} ~,\\
H =~& \text{sign}(y-\tau)\frac{h\sqrt{P(1-hP)(P-h)}}{1-hP-h'} ~,\\
Z =~& X -\frac{\E(h')}{\K(h')}Y ~,\\
A =~& \frac{\pi}{4\K(h)}\left(\frac{2Y}{\K(h')}-1 \right) ~,\\ 
T =~& \ln~\vartheta_2 \left( W_+~, -\pi\frac{\K(h')}{\K(h)}\right) -\ln~\vartheta_2 \left( W_-~, -\pi\frac{\K(h')}{\K(h)}\right) ~, \label{function-T}\\
W_\pm =~& \frac{\pi}{2\K(h)} \left[ \hI\pm \left( Y-\frac{\K(h')}{2} \right)\right] ~,\\
X =~& \text{sign}(y-\tau)\E(\vpd,h') ~,\\
Y =~& \text{sign}(y-\tau)\F(\vpd,h') ~,
\end{align}
\begin{align}
\vpd=~& \arcsin \left[ \frac{\sqrt{1-hP}}{h'} \right] ~,\\
P =~& \frac{\rac+2h\tau y}{(y+\tau)^2}
\end{align}
\end{subequations}
with
\begin{equation*}
\tau = \sqrt[4]{\mu^{-2}+1}
\end{equation*}
and $\vartheta_2$ being the Jacobian theta function defined by
\begin{equation*}
\vartheta_2(x;B)=\sum_{n=-\infty}^{\infty} \exp \left\{\left[\frac{1}{2}(2n+1)\right]^2 B + (2n+1)x \right\}.
\end{equation*}
The disc solution is physically relevant from the Newtonian limit given by $\mu \to 0$ to 
the ultra-relativistic limit, i.e.\ the black hole limit, given by the smallest positive value of $\mu$
for which $\cn(\hI,h')=0$. This ``upper limit'' of $\mu$ is called $\mu_0$, i.e.
\begin{equation}
0 < \mu \le \mu_0=4.6296618434743420427\dots
\end{equation}

\subsection{The black hole limit of the disc} \label{section-BH-limit-of-disk}
In the black hole limit, the source shrinks to the origin of the Weyl coordinate system, i.e.\ for the disc we have $\rho_0\to 0$. Therefore, it is appropriate to introduce two new pairs of dimensionless Weyl coordinates in addition to $x$ and $y$, one normalized with twice the angular velocity $\Omega$ 
and a second one normalized with the mass $M$ (note that $2\Omega M\to 1$ in the black hole limit, hence the two normalizations become identical in the limit):
\begin{subequations} \label{normalized-coord}
\begin{eqnarray}
\tilde{\rho}:=2\Omega\rho  ~,\quad & \dOz:=2\Omega\zeta &\quad\text{and}\\
\hat{\rho}:=\frac{\rho}{M}  ~,\quad & \zsM:=\dfrac{\zeta}{M} ~.&  
\end{eqnarray}
\end{subequations}
We will also use this ``tilde'' and ``hat'' notation to make some physical parameters dimensionless.
The logic of this notation is that a ``tilde'' introduces powers of $2\Omega$ such that the parameter becomes dimensionless
(mass $\mtz\equiv 2\Omega M$, angular momentum $\tilde{J}\equiv 4\Omega^2 J$, etc.), and the ``hat'' notation does
the same with powers of $M$ (mass $\hat{M}\equiv 1$, angular momentum $\hat{J}\equiv J/M^2$ etc.).

These normalized coordinates can be introduced in the Ernst potential of Eq.~(\ref{functions-of-mu-and-y})
by rewriting $y$ with the help of $2\Omega M=\mtz(\mu)$ and $\Omega\rho_0=\Omega_0(\mu)$ in one of the following two ways:
\begin{subequations} \label{coord-tranformation-for-disk}
\begin{align}
y~=~& \frac{\zeta}{\rho_0} \;=\; \frac{2\Omega M}{2\Omega\rho_0}\frac{\zeta}{M}
      \;=\; \frac{\mtz(\mu)}{2\Omega_0(\mu)}\zsM ~,\\
y~=~& \frac{\zeta}{\rho_0} \;=\; \frac{2\Omega\zeta}{2\Omega\rho_0} \;=\; \frac{\dOz}{2\Omega_0(\mu)}~.
\end{align}
\end{subequations}
The black hole limit of the disc is reached for $\mu \to \mu_0$, or equivalently for $\e^{V_0} \to 0$.
As discussed in the Introduction, there is a separation of spacetime in the limit. We are interested here in the ``outer world'' given by the extreme Kerr metric outside the horizon. Note that the horizon of the extreme Kerr metric is situated at $\rho=\zeta=0$. On the axis, the outer world corresponds to finite values of $\tilde{\zeta}$ or $\hat{\zeta}$, while the ``inner world'' is described by finite values of $y$. If we evaluate all the functions from Eq.~(\ref{f_of_disk})
in the black hole limit for $\hat{\zeta}>0$, we find:
\begin{equation}
N=1 ~, \quad Q_\pm=~1\pm \zsM ~.\label{lim_Q}
\end{equation}
This gives a potential which is identical 
to the Ernst potential of the extremal Kerr black hole on the (upper part of the) axis:
\begin{equation}
f(\mu_0;\zsM) = \frac{\zsM-1-\i}{\zsM+1-\i}  ~.\label{lim_f_of_disk}
\end{equation}
This potential can uniquely be extended off the axis and reads 
\begin{equation} \label{f-of-extreme-Kerr}
f=\frac{\hat{r}-1-\i\cos\theta}{\hat{r}+1-\i\cos\theta}~,
\end{equation}
where $\hat{r}:=\sqrt{\hat{\rho}^2+\hat{\zeta}^2}$ and $\tan\theta:=\hat{\rho}/\hat{\zeta}$.

\subsection{Leading order behaviour of the solution close to the black hole limit}\label{hair}
Taking the Ernst potential of any asymptotically flat solution at $\rho=0$, one can expand it at $\zeta\to +\infty$ in the following way:
\[ \frac{1-f(\zeta)}{1+f(\zeta)}= \sum_{n=0}^{\infty} \frac{m_n}{\zeta^{n+1}}~. \]
The coefficients $m_n$ are the same as those used in \cite{FHP89} to compute the multipole moments defined 
by Geroch \cite{Geroch70} and Hansen \cite{Hansen74}. The first two coefficients are  $m_0 = M$ and $m_1 = \i\,J$.
Let us rewrite the coefficients in a normalized and dimensionless form:
\[ m^*_n:=\frac{m_n}{k_n} \quad\text{with}\quad k_n:=M\left(\frac{\i J}{M}\right)^n=\frac{m_1^n}{m_0^{n-1}}~. \]
Note that the $k_n$ correspond exactly to the coefficients $m_n$ of a Kerr solution with mass $M$ and angular momentum $J$, meaning that $m^*_n = 1$ for black holes.

It turns out that the following relations hold in the black hole limit: 
\begin{subequations}
\begin{align}
\left.\frac{\d\, m^*_n}{\d\,\mu  }\right |_{\mu=\mu_0}&=0~,  \\
\left.\frac{\d^2 m^*_n}{\d\,\mu^2}\right |_{\mu=\mu_0}&=0 
\end{align}
\end{subequations}
or, equivalently,
\begin{subequations}\label{properties}
\begin{align}
\left.\frac{\d\, m^*_n}{\d\,\e^{V_0}}\right |_{V_0\to-\infty}&=0~, \label{property-1} \\
\left.\frac{\d^2 m^*_n}{\d\,(\e^{V_0})^2}\right |_{V_0\to-\infty}&=0~. \label{property-2}
\end{align}
\end{subequations}
Actually, numerical evidence suggests strongly that \eqref{properties} holds for a large class of solutions (not only the disc) with parametric transition 
to a black hole\footnote{It was shown in  \cite{m06} that $\e^{V_0}\to 0$ is necessary and sufficient for reaching a black hole limit. An equivalent condition is the relation \eqref{BHcond}.}. Indeed, Eq.~\eqref{property-1} was verified in \cite{LPA07} (in a different but equivalent form) for several sequences of numerical solutions describing rotating fluid rings reaching the black hole limit.
We have checked that the second property \eqref{property-2} holds for these sequences as well. Note that the relations \eqref{properties} also hold for the normalized Geroch-Hansen multipole moments ($m^*_n$ replaced by the Geroch-Hansen moments normalized with the $k_n$). This follows directly from the structure of the expressions given in \cite{FHP89}.  

It can easily be concluded that the Ernst potential of any fluid configuration having a black hole limit ($\e^{V_0}\to 0$) and fulfilling the conditions \eqref{properties} can be decomposed into an Ernst potential $f_{\rm Kerr}$ of the Kerr solution with the same mass and angular momentum
and a residual potential $R$  of order $\e^{3V_0}$:
\begin{subequations} \label{decomposition-of-f}
\begin{align}
f(M,J;\zeta)   =~& f_{\rm Kerr}(M,J;\zeta) \;+\; R(M,J;\zeta)~, \\
f_{\rm Kerr}(M,J;\zeta) =~& \frac{M(\zeta-M)-\i J}{M(\zeta+M)-\i J}~, \\
  R(M,J;\zeta) =~& {\cal O}(\e^{3V_0})~.
\end{align}
\end{subequations}
Note that $f_{\rm Kerr}$ is not restricted to $J\le M^2$ as it is for black holes.
Indeed, rings and discs have $J\ge M^2$, thus $f_{\rm Kerr}$ is given here by the axis potential of  
a ``hyperextreme'' Kerr solution. In the above equations one may take $M$ as one parameter and consider $J$ as a function of $M$ and of the second parameter $\e^{V_0}$, i.e.\ $J=J(M,\e^{V_0})$. For the disc solution, $J/M^2$ is a function of $\e^{V_0}$ alone. At the black hole limit, $\e^{V_0}=0$, $J=M^2$ and
$R$ vanishes. Comparing Eqs~\eqref{decomposition-of-f} 
with Eq.~\eqref{f-at-infinity} one can see that $R$ must also decay rapidly in the far field, $R = {\cal O}(\zeta^{-3})$, corresponding to the fact that $f_{\rm Kerr}$ already contributes the correct first two multipole moments. 
Eqs~\eqref{decomposition-of-f} can be extended off the axis with $f_{\rm Kerr}$ as the Ernst potential of the general Kerr solution. While $f$ and $f_{\rm Kerr}$ are then both solutions to the Ernst equation, $R$ is of course not -- due to the non-linearity.

\section{Taylor series at the black hole limit}
Our aim is to expand the Ernst potential $f(\mu;\hat{\zeta})$ or $f(\mu;\tilde{\zeta})$ of the disc solution at the black hole limit ($\mu\to\mu_0$) in terms of powers of $\muK$. Such an expansion can then easily be transformed to an expansion in terms of the more universal parameter $\e^{V_0}$.

To find an explicit form of the Taylor series, we make use of the computer algebra systems
Maple and Mathematica and we divide the work in two steps. 
First, we write down the series of all functions that depend on $\mu$ alone. Then, we do
the same for the remaining functions which depend on $\mu$ and $\hat{\zeta}$ or $\tilde{\zeta}$. 
\subsection{Series of functions of $\mu$}
\begin{figure}
\begin{center}
\includegraphics[scale=0.6]{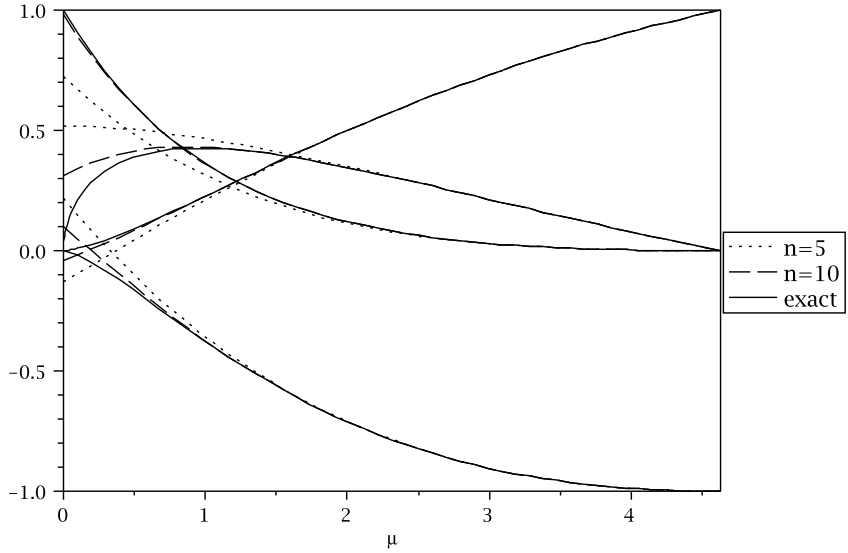}
\caption{Four exact functions of $\mu$ compared to their Taylor approximations with orders $n=5$ and $10$. On the right side of the plot, the functions are from the top to the bottom:
$2\Omega M$, $\Omega\rho_0$, $\e^{2V_0}$ and $b_0$.\label{figs-four-functions-and-series}}
\end{center}
\end{figure}
The set of functions that depend on $\mu$ alone is listed in Eqs~\eqref{functions-of-mu}. 
For each of these functions, the resulting derivatives of $\mu$ in the limit $\mu \to \mu_0$ give pure numbers.

The direct computation of a series for $\hI$ with the form given previously is impracticable because zeros come out for some denominators. To avoid this pathology, we
rewrite the function with the substitution $x=\mu \sin^2\phi$. This also has the effect
of removing the $\mu$-dependence of the upper limit of integration:
\begin{equation} \label{Ic-substitution}
\hI(\mu)
= \frac{\sqrt[4]{1+\mu^2}}{\pi}\int\limits^{\mu}_0 \frac{g(x)}{\sqrt{\mu-x}}\d x
= \frac{2\sqrt[4]{\mu^2+\mu^4}}{\pi}\int\limits^{\pi/2}_0 g(\mu \sin^2\phi) \sin\phi \,\d\phi.
\end{equation}
The series reads
\begin{align}\label{series-of-Ic}
\hI(\mu)=~& 1.5752 +0.24046\,\muK -0.017245\,\muK^2 \nn &
 +0.0017270\,\muK^3 + {\cal O}[\muK^4] ~,
\end{align}
where decimal numbers are truncated after 5 significant digits (we omit the ellipsis after each number from here on).
Similarly, the derivatives of all other functions of $\mu$ that we need for the Ernst potential can be calculated recursively. 
Thus we can already present Taylor series of a few relevant parameter functions 
such as the Ernst potential at the centre of the disc, given by $f(\rho=0,\zeta=0^+)\equiv f_0= \e^{2V_0}+\i b_0$,
as well as the dimensionless products $\Omega\rho_0$ and $2\Omega M$. These functions need to be expanded with high numerical precision since they enter into the remaining series expansions.
\begin{table}
 \centerline{\begin{tabular}{c|rrrr}
  $j$ & $\e^{2V_0}$ & $b_0$ & $\Omega\rho_0$ & $2\Omega M$ \\ \hline 
   0  &  0                         & -1                         &  0                         &  1 \\
   1  &  0                         &  0                         & -1.1979704$\times 10^{-1}$ &  1.2563637$\times 10^{-1}$ \\
   2  &  6.1997318$\times 10^{-3}$ &  2.8702661$\times 10^{-2}$ &  8.2373333$\times 10^{-3}$ & -2.2207483$\times 10^{-2}$ \\
   3  & -2.1917290$\times 10^{-3}$ & -3.9472326$\times 10^{-3}$ &  4.3533289$\times 10^{-4}$ &  1.1246071$\times 10^{-3}$ \\
   4  &  4.5766410$\times 10^{-4}$ &  3.5824068$\times 10^{-4}$ & -1.9230828$\times 10^{-4}$ &  1.7087311$\times 10^{-5}$ \\
   5  & -7.5851829$\times 10^{-5}$ & -2.0388433$\times 10^{-5}$ &  3.8239383$\times 10^{-5}$ & -1.4784817$\times 10^{-5}$ \\
   6  &  1.1139165$\times 10^{-5}$ & -9.6924305$\times 10^{-7}$ & -6.5882003$\times 10^{-6}$ &  3.0205335$\times 10^{-6}$ \\
   7  & -1.5243204$\times 10^{-6}$ &  6.0947891$\times 10^{-7}$ &  1.0947125$\times 10^{-6}$ & -5.0848591$\times 10^{-7}$ \\
   8  &  1.9941471$\times 10^{-7}$ & -1.4423676$\times 10^{-7}$ & -1.8180382$\times 10^{-7}$ &  8.0708022$\times 10^{-8}$ \\
   9  & -2.5296376$\times 10^{-8}$ &  2.7737508$\times 10^{-8}$ &  3.0627438$\times 10^{-8}$ & -1.2616141$\times 10^{-8}$ \\
  10  &  3.1381072$\times 10^{-9}$ & -4.9011518$\times 10^{-9}$ & -5.2656087$\times 10^{-9}$ &  1.9805535$\times 10^{-9}$
  \end{tabular}}
  \caption{First coefficients $a_j$ of the expansions defined in Eq.\eqref{series-of-func-of-mu}
           for the functions $\e^{2V_0}$, $b_0$, $\Omega\rho_0$ and $2\Omega M$. 
           \label{coeff-of-func-of-mu}}
\end{table}

\begin{figure}
\subfloat[$\Big|\e^{2V_0}-{\cal A}_n(\e^{2V_0})\Big|$]{\includegraphics[scale=0.3]{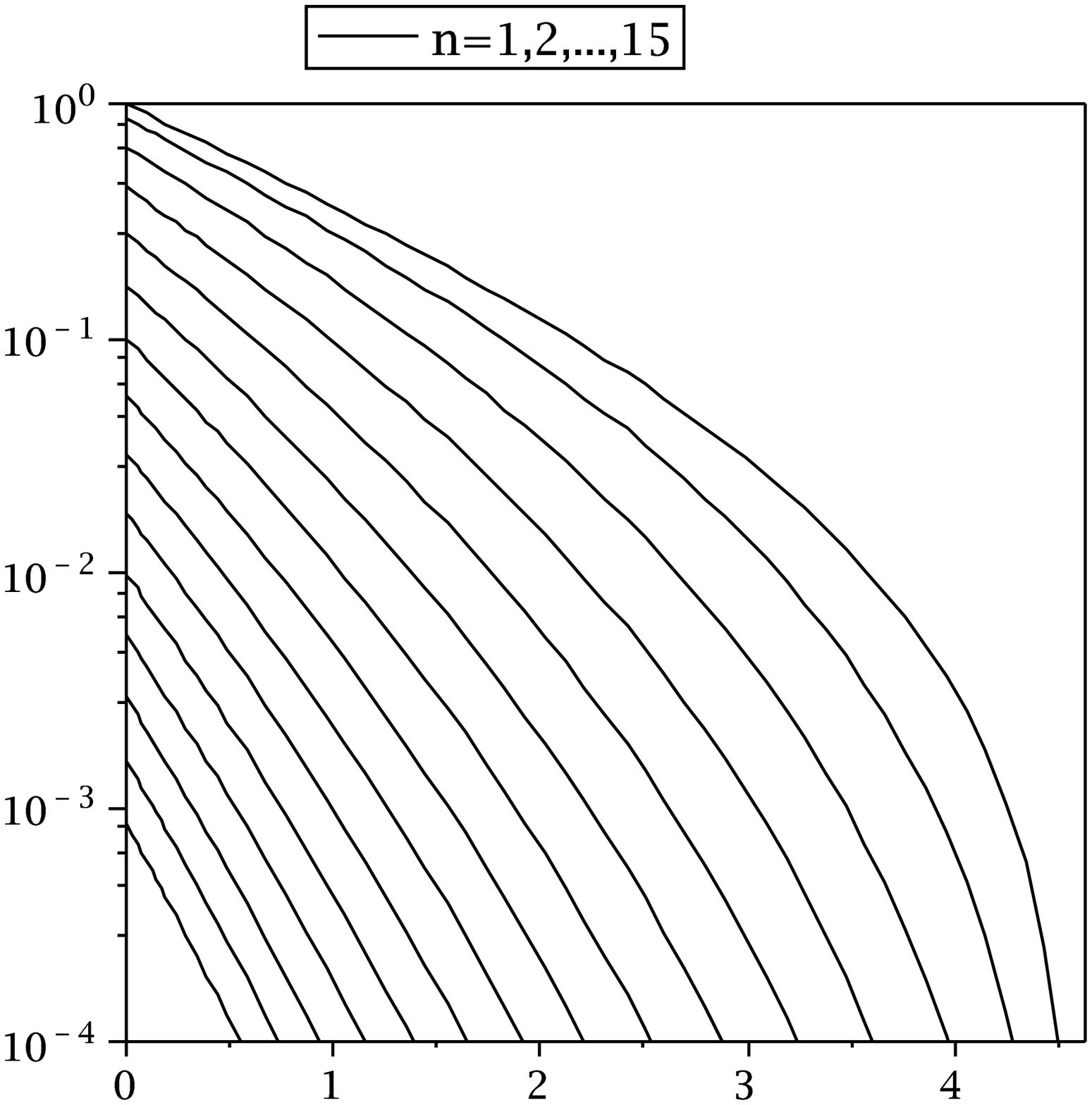}}
\subfloat[$\Big|b_0-{\cal A}_n(b_0)\Big|$]{\includegraphics[scale=0.3]{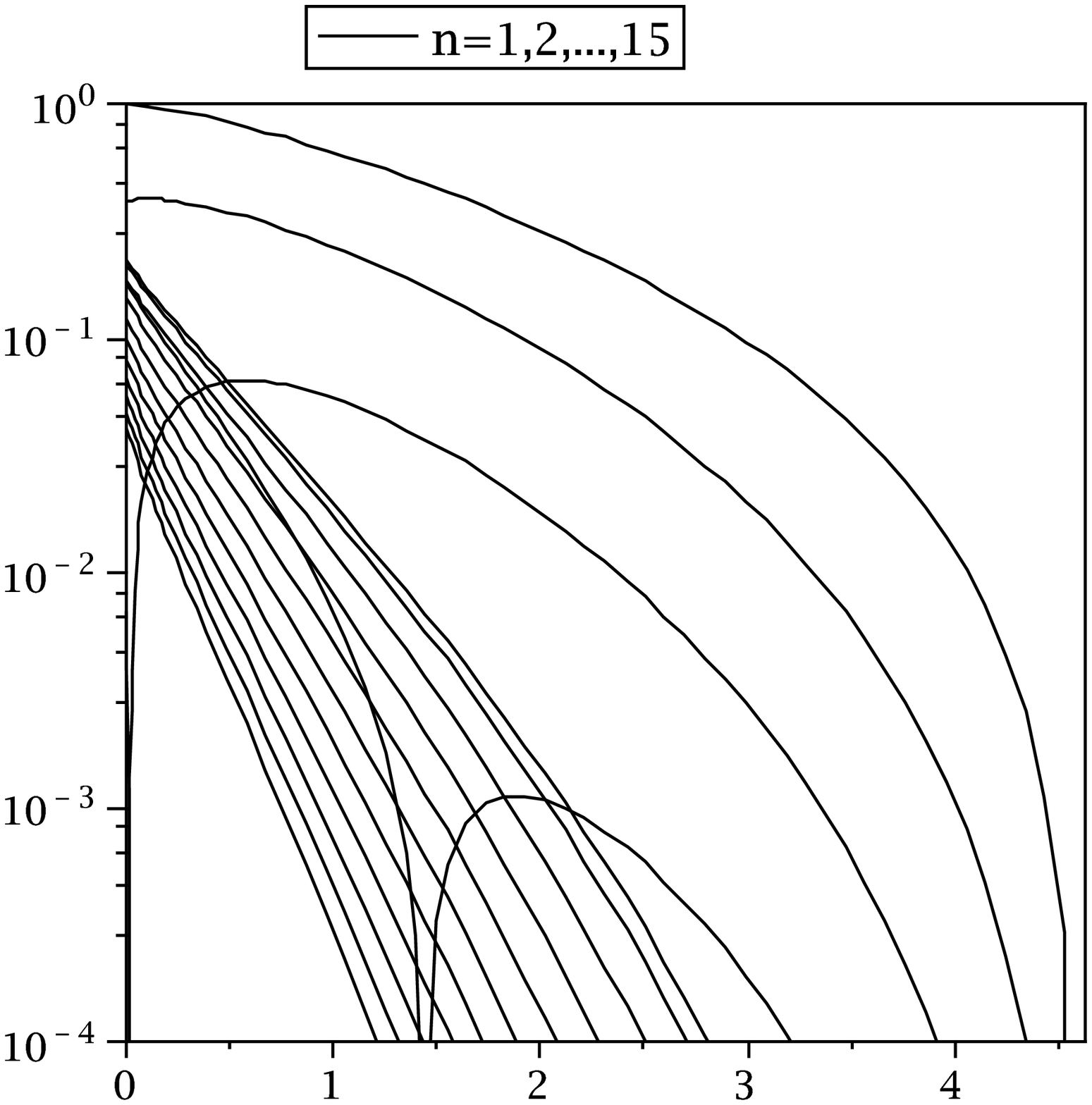}} \\ 
\subfloat[$\Big|\Omega \rho_0-{\cal A}_n(\Omega \rho_0)\Big|$]{\includegraphics[scale=0.3]{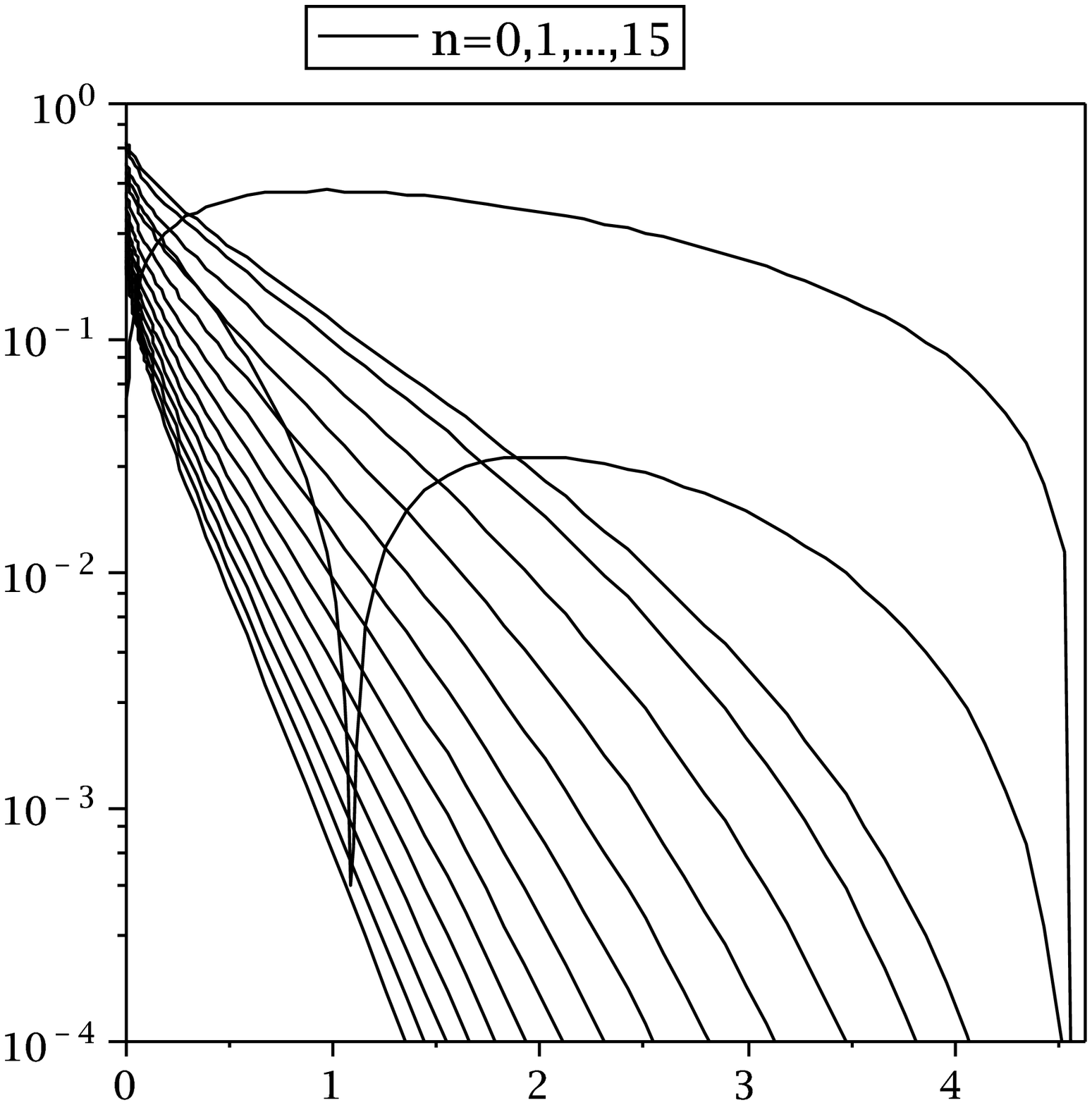}}
\subfloat[$\Big|2\Omega M-{\cal A}_n(2\Omega M)\Big|$]{\includegraphics[scale=0.3]{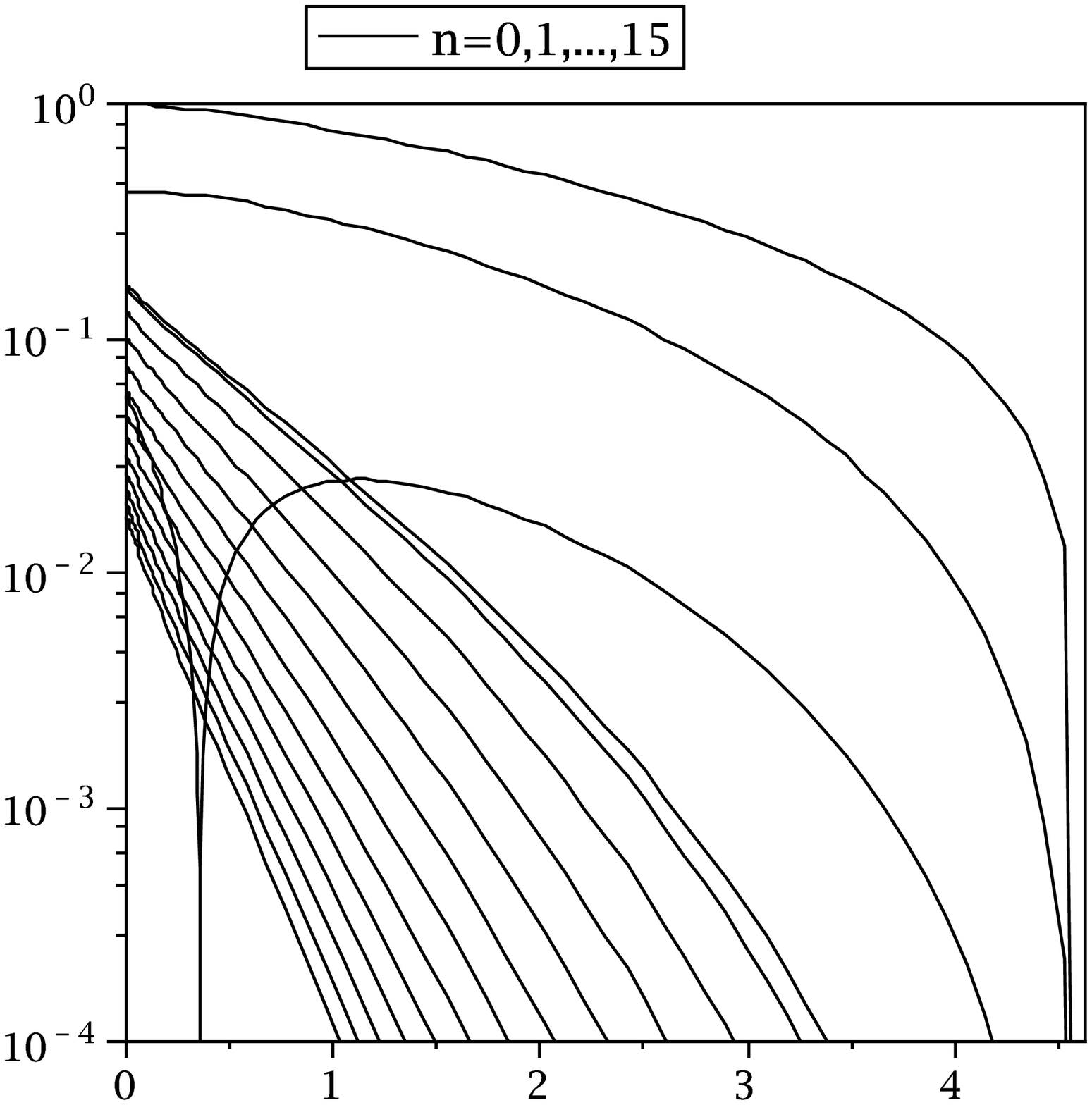}}
\caption{These figures show the deviations of the $n$-th order Taylor approximations ${\cal A}_n$
from their respective exact functions. The Taylor series are truncated to order $n=1\dots 15$ 
for the functions $\e^{2V_0}$  and $b_0$, and to $n=0\dots 15$ for $\Omega \rho_0$ and $2\Omega M$.
The lines are ordered from the smallest $n$ to the largest when one follows the abcisssa from right to left.
The parameter $\mu$ is shown on the abscissae and the ordinates show on a logarithmic scale 
the absolute value of the deviations.
The plunges to zero for a few curves only mean that the corresponding polynomials ${\cal A}_n$
intersect the exact function at that point.}\label{figs-series-of-mu}
\end{figure}

Let us now introduce the following notation for the $n$-th order Taylor approximation 
${\cal A}_n$ of a function of $\mu$ near the black hole limit:
\begin{align} \label{series-of-func-of-mu}
{\cal A}_n\left(b_0(\mu)\right) &= \sum_{j=0}^n a_j \muK^j~, 
\end{align}
where $b_0(\mu)$ is used as an example.
For the short list of functions that we have introduced, we show their first expansion coefficients $a_j$ in Table~\ref{coeff-of-func-of-mu}. The same functions are also plotted in 
Fig.~\ref{figs-four-functions-and-series} with their respective Taylor approximations of order $n=5$ and $n=10$.
This figure allows us to assess the quality of the Taylor series.
Approximations with $n=5$ and $n=10$ are indistinguishable from the exact function on the plot for $\mu>2$, while polynomials of higher orders ($n=10$ versus $n=5$ on the plot) improve the approximation near the Newtonian limit ($\mu \to 0$) reasonably well.
In Fig.~\ref{figs-series-of-mu}, one can see that the Taylor series seem to converge to their respective exact functions from the black hole limit ($\mu\to\mu_0$) all the way down to the Newtonian limit ($\mu\to 0$).
The convergence is readily seen in the figure with $\e^{2V_0}$, while for the three other functions, one must take into account the logarithmic scale to appreciate it.
Since the function $\e^{2V_0}$ is related to the redshift parameter
as given in Eq.~\eqref{redshift}, we can calculate how wrong the redshift becomes from the approximations. In the Newtonian limit, where the Taylor approximations have the greatest deviations, the redshifts, approximated by $Z_0^{(n)}=[{\cal A}_n(\e^{2V_0})]^{-1/2}-1$,  become $Z_0=0.17691,~9.0567\times 10^{-3},~4.2302\times 10^{-4}$  for $n=5,~10,~15$ respectively, while of course $Z_0\to 0$ is the correct result.

\subsection{Series of functions of $\mu$ and $\dOz$} \label{section-series-of-function-mu-dOz}

The next step to obtain a series of the disc's axis potential is to expand the remaining terms which depend on both $\mu$ and $\tilde{\zeta}$. To this end, we introduce the normalized coordinate $\dOz\equiv 2\Omega\zeta$ in every function given in Eqs~\eqref{functions-of-mu-and-y}.
By computing series at $\mu=\mu_0$ for $L(\mu,\dOz)$, 
$S(\mu,\dOz)$, $H(\mu,\dOz)$, $Z(\mu,\dOz)$, $A(\mu,\dOz)$ and
$T(\mu,\dOz)$, we can then determine the series of $N(\mu,\dOz)$, which can be combined
with the series of $Q_\pm (\mu,\dOz)$ to obtain the Ernst potential of the disc on the axis.

For the function $L(\mu,\dOz)$, it is helpful to carry out the substitution
$x=\mu\sin^2 \phi$, as in Eq.~\eqref{Ic-substitution}, in order to calculate it numerically:
\begin{align*}
&L(\mu,\dOz)= \\
&\frac{4\Orz\dOz}{\pi\sqrt{(\dOz^2+4\Orz^2)^2\mu^2+16\Orz^4}} 
\int\limits_0^{\pi/2} \frac{(\dOz^2+4\Orz^2)\mu^2\sin^2\phi+4\Orz^2}{\dOz^2+4\Orz^2\cos^2\phi} 
g(\mu\sin^2\phi) \sin\phi\, \d \phi,
\end{align*}
where $\Orz$ is a function of $\mu$ already expanded before. Once the integrand is expanded
in a series in $\muK$, the integral becomes easier to perform on each individual term of the expansion.
This series contains polynomials of $1/\dOz$ with odd exponents and reads
\begin{align}\label{series-of-R-dOz}
L(\mu,\dOz)=&
- \frac{0.24697}{\dOz}\muK
- \frac{0.015102}{\dOz}\muK^2 \nn &
+\left( \frac{0.0063269}{\dOz} + \frac{0.0021138}{\dOz^3} \right) \muK^3
+{\cal O}[\muK^4]~,
\end{align}
where numbers are again truncated after five significant digits. The next functions to expand,
$S(\mu,\dOz)$ and $P(\mu,\dOz)$, do not require anything other than a direct computation
of the series from the computer. The series for $S$ reads
\begin{align}\label{series-of-S-dOz}
S(\mu,\dOz)=~& 0.053082 - 0.011080\,\muK \\ & 
+ \left( 0.0022735 - \frac{0.0061646}{\dOz^2} \right) \muK^2 \nn &
+ \left( -4.5822\times 10^{-4} +\frac{0.0021646}{\dOz^2} \right) \muK^3 
+ {\cal O}[\muK^4] ~. \nonumber
\end{align}
The function $H(\mu,\dOz)$ is not difficult to expand by calling $P(\mu,\dOz)$ 
where needed. Note that the sign from the factor ``$\text{sign}(y-\tau)$'' has to be chosen to be positive since $y\to\infty$ for finite $\tilde{\zeta}$ at the black hole limit.
Of course, our series will diverge in a certain region near the disc. This is related to the previously discussed ``separation of spacetime'' as $\mu\to\mu_0$. The result for $H$ is
\begin{align}\label{series-of-V-dOz}
H(\mu,\dOz)=~& -0.055908  +\left( 0.012264 +\frac{0.0013618}{\dOz} \right)\muK \nn & 
+\left( -0.0026412 -\frac{6.6713\times 10^{-4}}{\dOz} +\frac{0.0064947}{\dOz^2} \right)\muK^2 \nn &
+\left( 5.5869\times 10^{-4} +\frac{2.1410\times 10^{-4}}{\dOz} -\frac{0.0023506}{\dOz^2} \right. 
\nn & \left. 
-\frac{2.3723\times 10^{-4}}{\dOz^3} \right)\muK^3
+ {\cal O}[\muK^4] ~.
\end{align}
The remaining functions $Z(\mu,\dOz)$, $A(\mu,\dOz)$ and $T(\mu,\dOz)$ all
depend on the elliptic integrals given by $X(\mu,\dOz)$ and $Y(\mu,\dOz)$.
Since we restrict ourselves from now on to the above mentioned ``plus sign'', the latter functions become simply $X=\E(\vpd,h')$ and $Y=\F(\vpd,h')$. A good strategy to produce series of these functions is given by expanding the right hand sides of
\begin{align*}
\vpd                      = \am(Y,h') &= \arcsin \left[ \frac{\sqrt{1-hP}}{h'} \right] ~,\\
\sin(\vpd)                = \sn(Y,h') &= \frac{\sqrt{1-hP}}{h'} ~,\\
\cos(\vpd)                = \cn(Y,h') &= \frac{\sqrt{h\,(P-h)}}{h'} ~,\\
\sqrt{1-h'^2\sin^2(\vpd)} = \dn(Y,h') &= \sqrt{hP} ~.
\end{align*}
Then, the Taylor series of $\F(\vpd,h')$ and $\E(\vpd,h')$ can be computed as series
containing derivatives of the Jacobian elliptic functions.
Each time that such derivatives need to be evaluated, 
the answer can easily be picked up in the four series obtained from above.
By properly combining the series of $\F(\vpd,h')$, $\E(\vpd,h')$, $\K(h)$, $\K(h')$ and $\E(h')$,
we obtain:
\begin{align}
Z(\mu,\dOz)=~&  0.0028257 +\left( -0.0011830 +\frac{1.9350\times 10^{-6}}{\dOz} \right)\muK \nn &
+\left( 3.6763\times 10^{-4} -\frac{1.7626\times 10^{-6}}{\dOz} 
-\frac{3.3002\times 10^{-4}}{\dOz^2} \right)\muK^2 \nn &
+\left( -1.0047\times 10^{-4} +\frac{9.5848\times 10^{-7}}{\dOz} 
+\frac{1.8595\times 10^{-4}}{\dOz^2} \right. \nn & - \left. \frac{3.3836\times 10^{-7}}{\dOz^3} \right)\muK^3
+ {\cal O}[\muK^4] ~,\label{series-of-Z-dOz}
\end{align}
\begin{align} \label{series-of-U-dOz} 
A(\mu,\dOz)=~&  -\frac{0.066452}{\dOz}\muK + \frac{0.0085785}{\dOz}\muK^2 ~\\ &
+\left( -\frac{7.1026\times 10^{-4}}{\dOz} + \frac{0.0012715}{\dOz^3} \right)\muK^3
+ {\cal O}[\muK^4] ~, \nonumber
\end{align}
\begin{align}
T(\mu,\dOz)=~& -\frac{0.20935}{\dOz}\muK -\frac{0.0042719}{\dOz}\muK^2 \nn &
+\left( \frac{0.0038072}{\dOz} +\frac{0.0040061}{\dOz^3} \right)\muK^3
+ {\cal O}[\muK^4] ~. \label{series-of-T-dOz}
\end{align}
Note that we obtained the series of $T$ by the calculation of a series of $\d T/\d \mu$ using several relations for derivatives of the theta function $\vartheta_2$ and subsequent integration. 
At this stage, the remaining computations are straightfoward. 
The series of $N(\mu,\dOz)$ is obtained by combining together the series from
Eqs~\eqref{series-of-Ic} and (\ref{series-of-R-dOz}\,--\,\ref{series-of-T-dOz}):
\begin{align}\label{series-of-N-dOz}
N(\mu,\dOz)=~& 1 -\frac{0.25127}{\dOz}\muK 
+\left( -\frac{0.012990}{\dOz} +\frac{0.031568}{\dOz^2} \right)\muK^2 \nn &
+\left( \frac{0.0056452}{\dOz} +\frac{0.0032641}{\dOz^2} +\frac{2.1819\times 10^{-4}}{\dOz^3} \right)\muK^3 \nn &
+ {\cal O}[\muK^4] ~.
\end{align}
And the expansions of $\Orz(\mu)$ and $b_0(\mu)$ are needed for $Q_\pm(\mu,\dOz)$,
which gives:
\begin{equation}\label{series-of-Q-dOz}
Q_\pm(\mu,\dOz)= 1 \pm \dOz -0.028702 \,\muK^2 +0.0039472\,\muK^3 + {\cal O}[\muK^4] ~.
\end{equation}

\subsection{Series of the Ernst potential of the disc}
\begin{figure}
\begin{center}
\subfloat[$\e^{2U}$ at $\dOz=10$]{\includegraphics[scale=0.26]{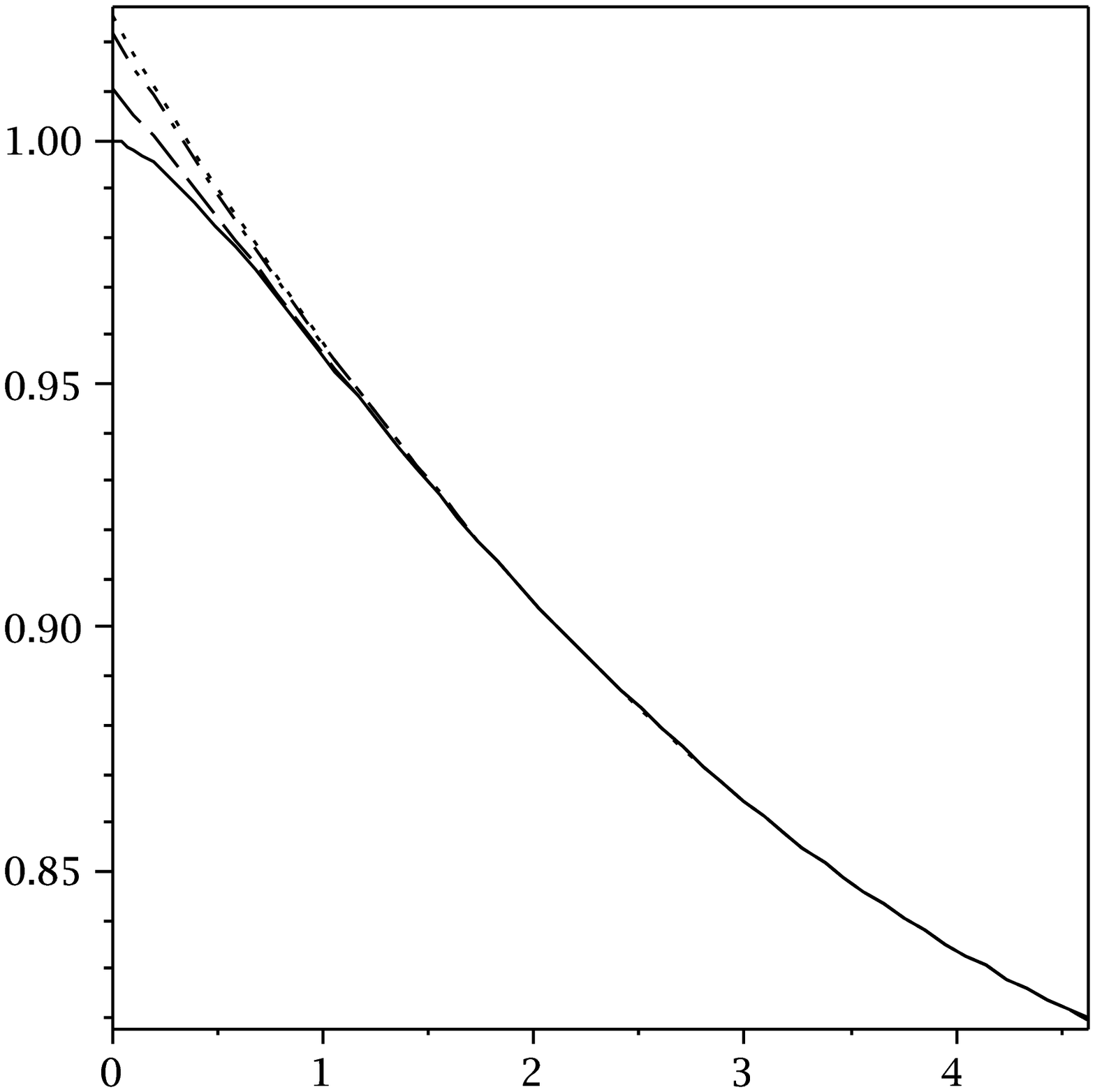}}
\subfloat[$b$ at $\dOz=10$]{\includegraphics[scale=0.26]{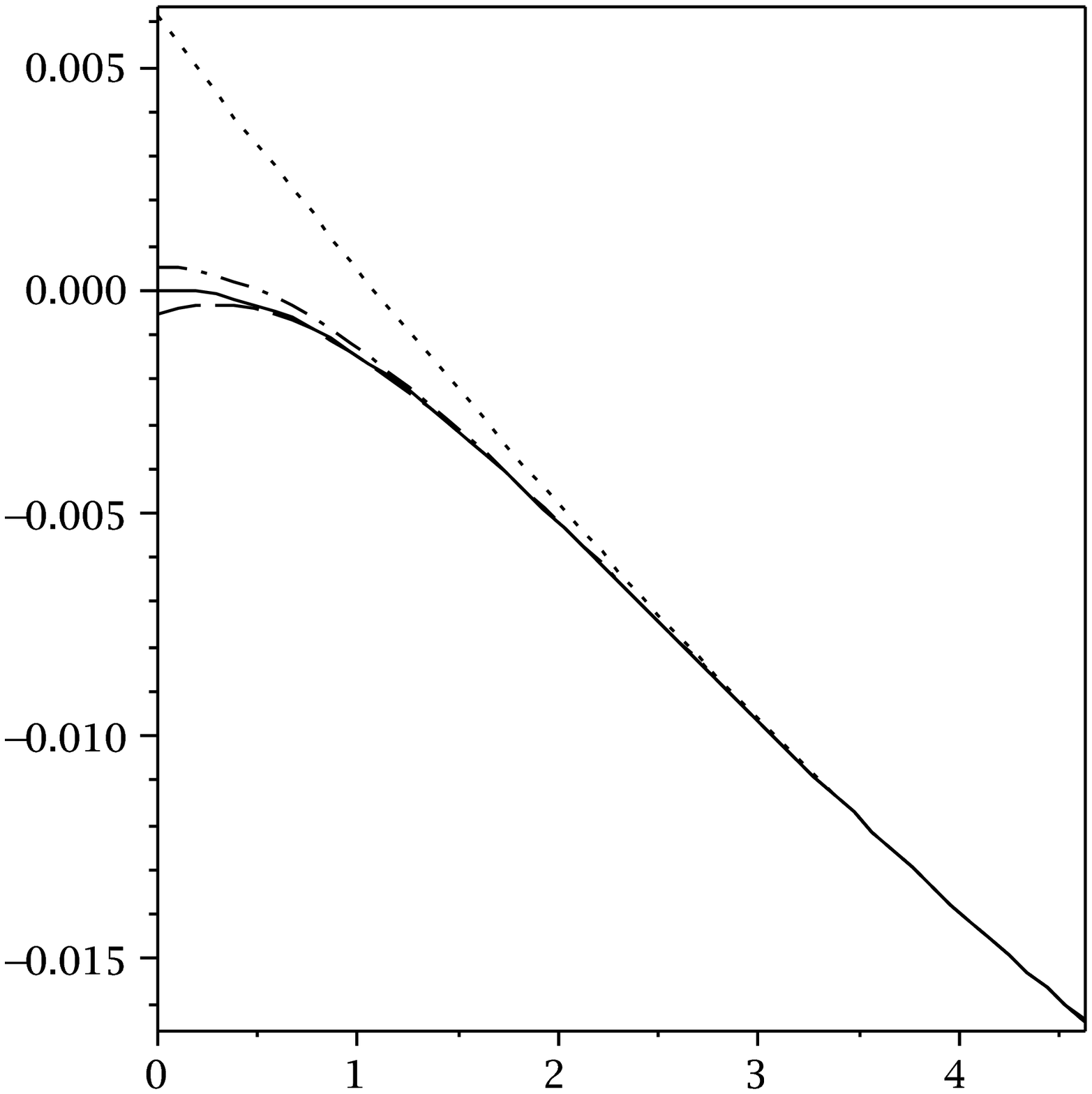}} \\
\subfloat[$\e^{2U}$ at $\dOz=2$]{\includegraphics[scale=0.26]{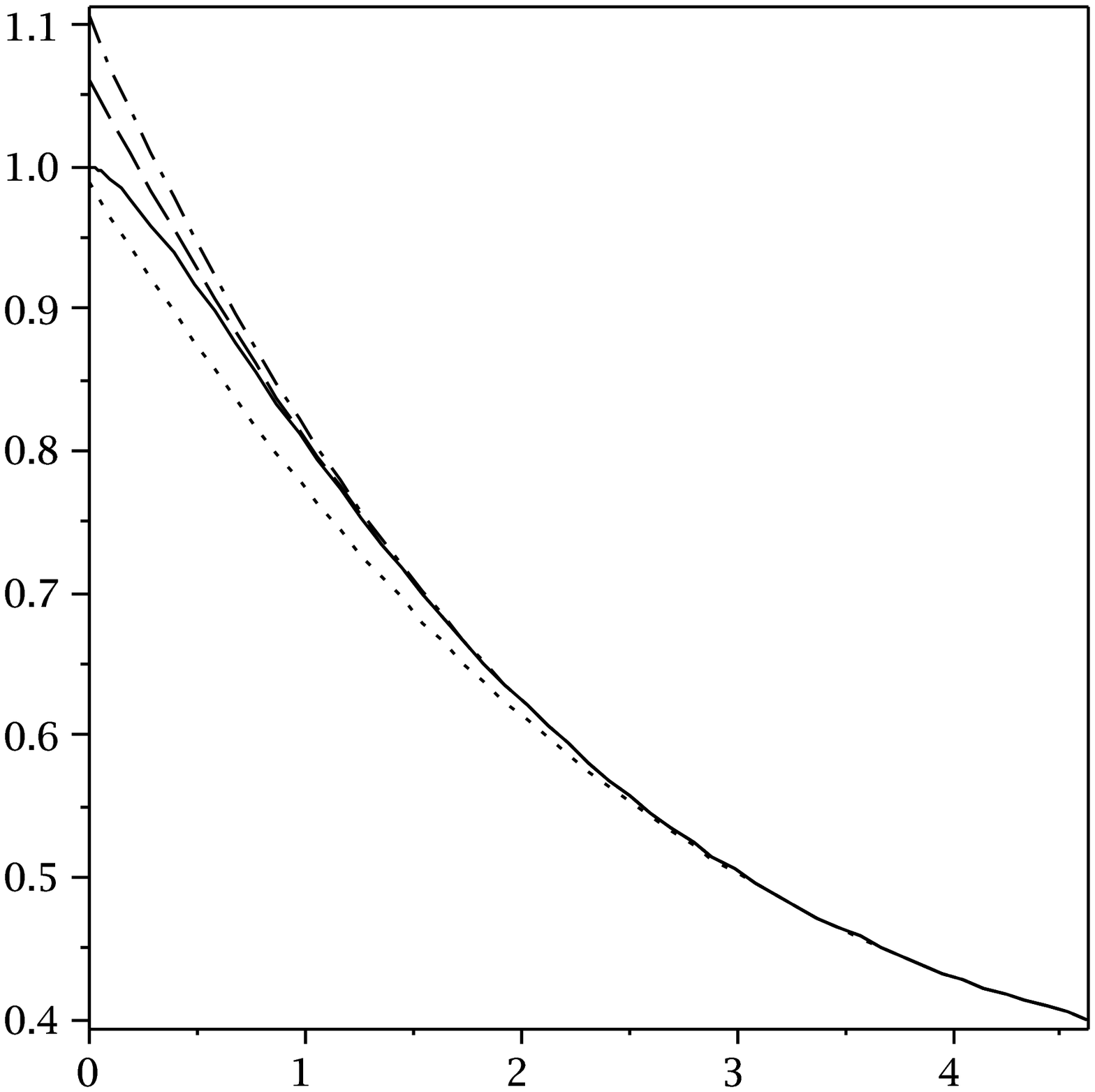}}
\subfloat[$b$ at $\dOz=2$]{\includegraphics[scale=0.26]{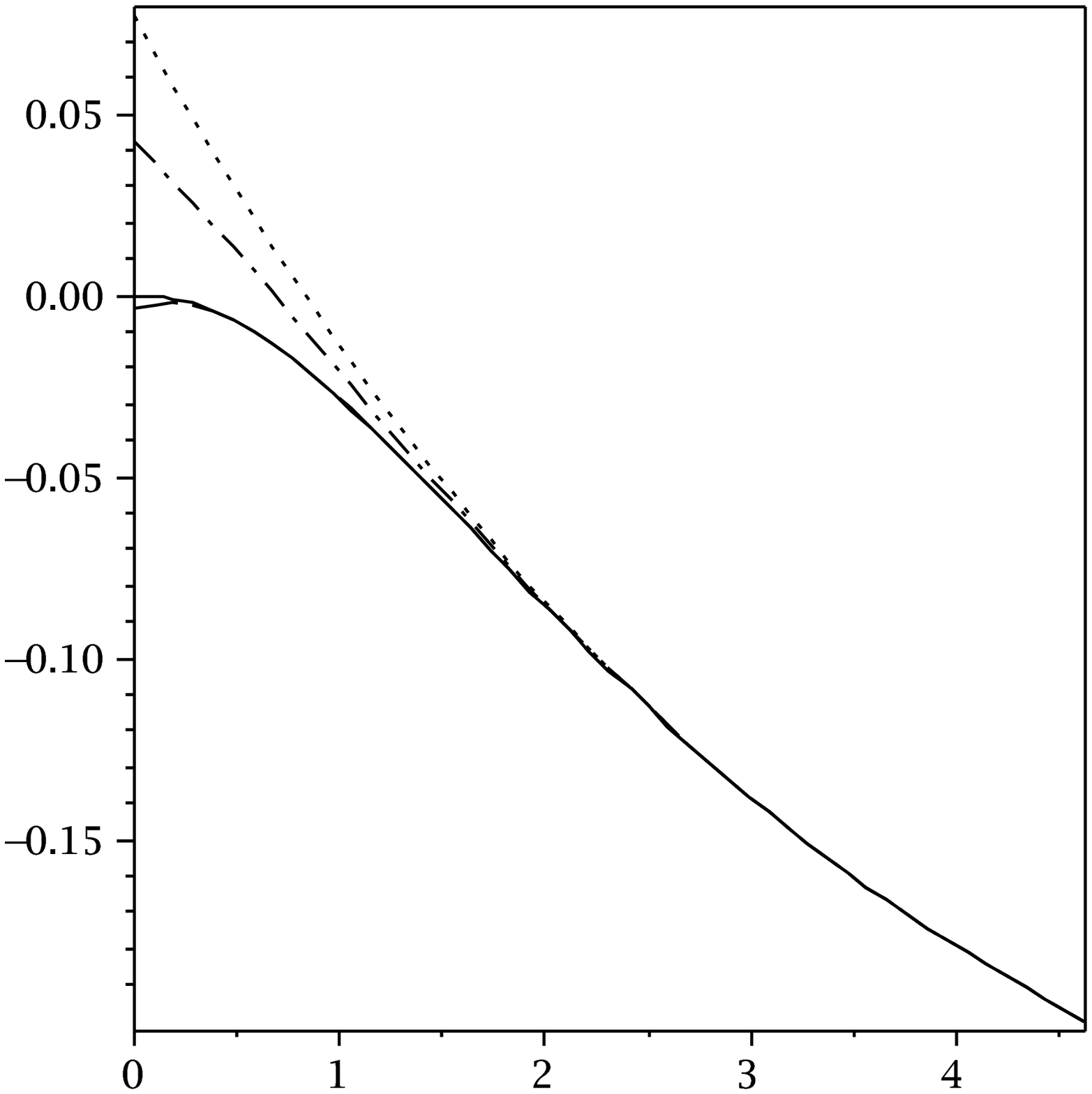}} \\
\subfloat[$\e^{2U}$ at $\dOz=0.5$]{\includegraphics[scale=0.26]{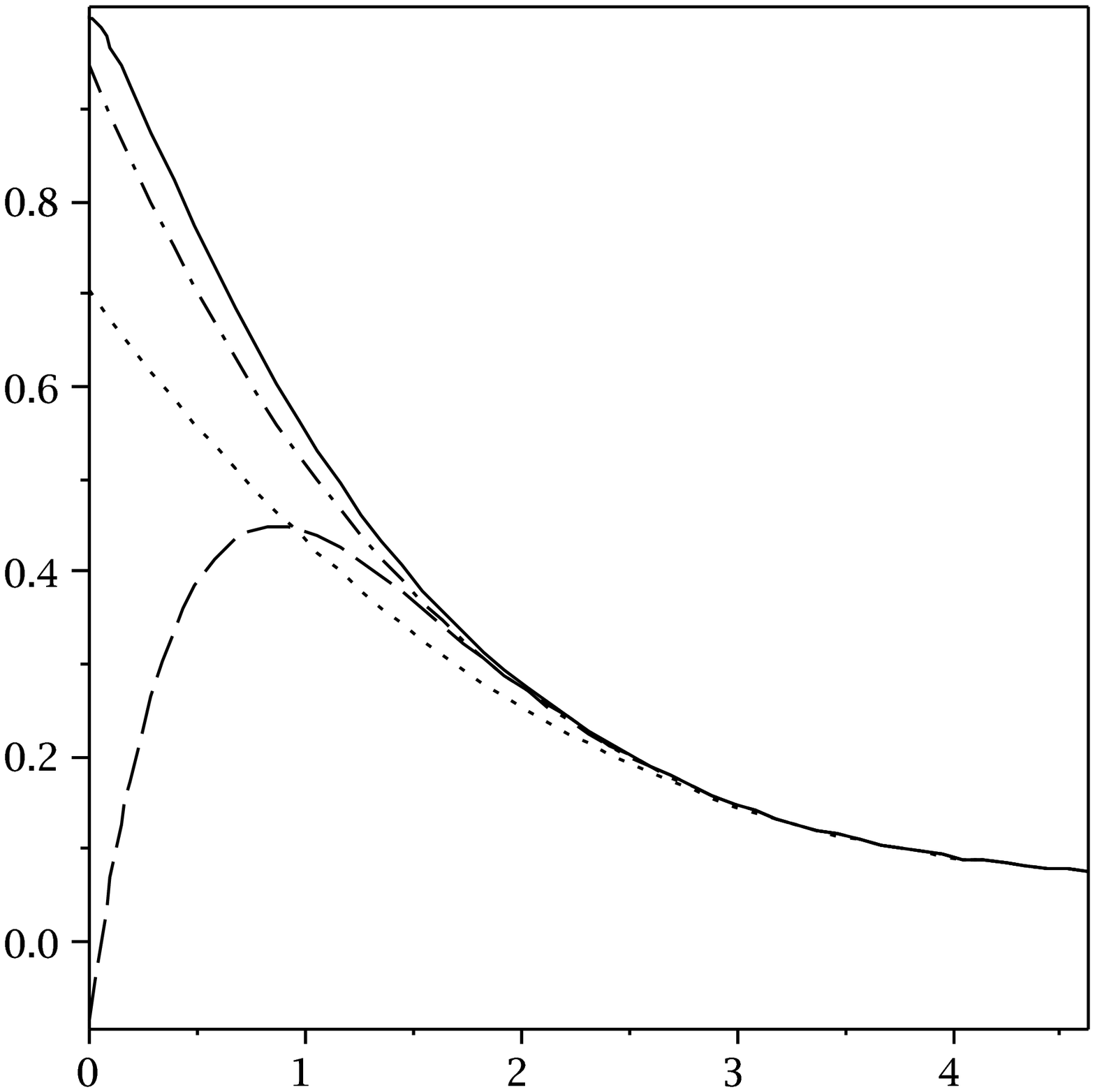}}
\subfloat[$b$ at $\dOz=0.5$]{\includegraphics[scale=0.26]{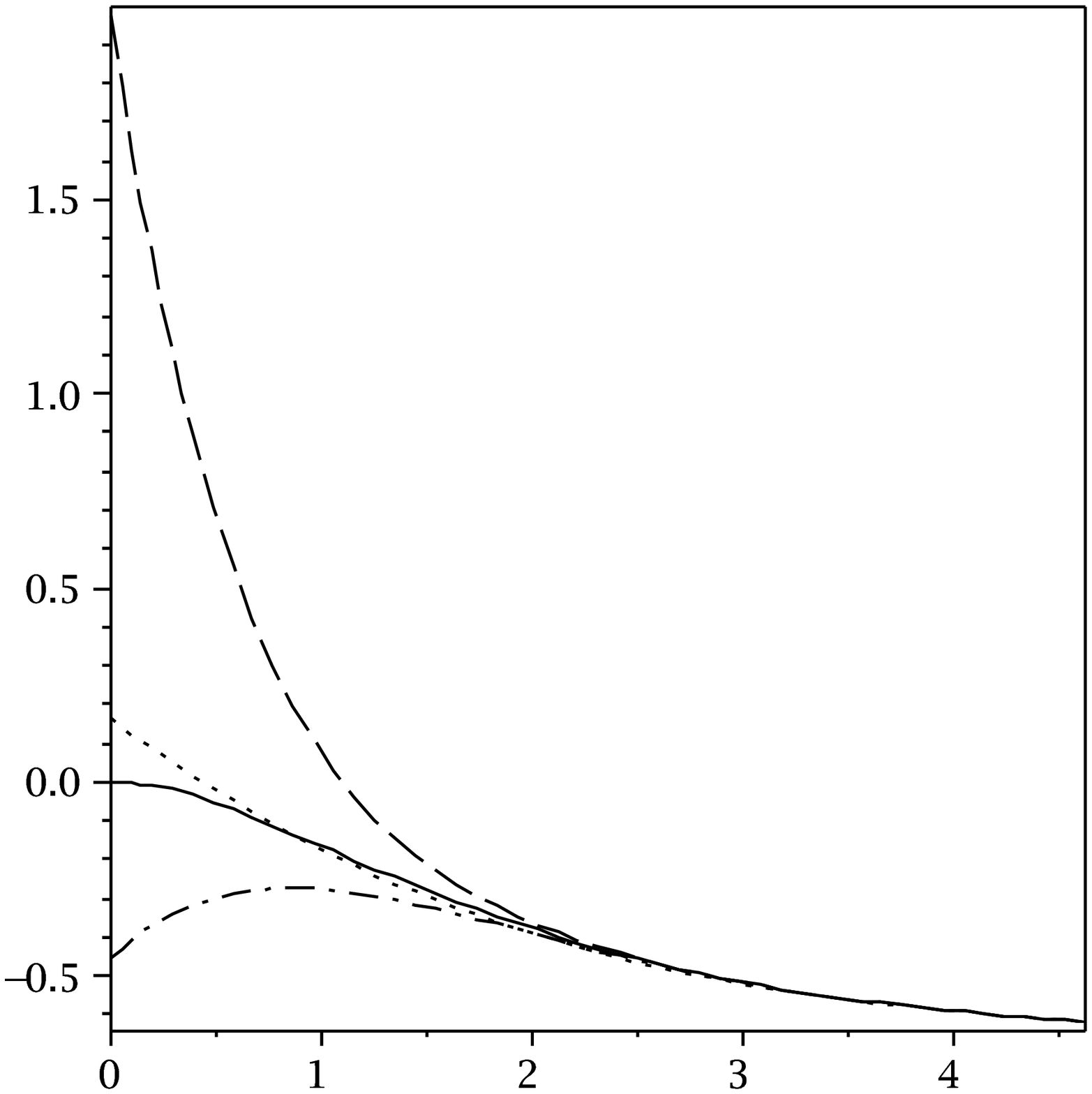}}
\caption{Real and imaginary parts of the Ernst potential and their respective Taylor approximation
as function of $\mu$ for $\dOz=$10, 2 and 0.5. The solid lines are from the exact Ernst potential, 
while the dotted, dash-dotted and dashed lines are Taylor polynomials with orders $n=3,6,9$ respectively.
\label{f_series_zeta_tilde}}
\end{center}
\end{figure}
Both real series \eqref{series-of-N-dOz} and \eqref{series-of-Q-dOz} can now be combined
into the complex Ernst potential given by Eq.~\eqref{f_of_disk}. The series
takes the following form:
\begin{align}\label{series-of-f-disk-tilde}
f(\mu;\dOz) =~& \frac{\dOz-1-\i}{\dOz+1-\i} - \frac{0.25127\,\dOz}{(\dOz+1-\i)^2}\muK \nn
& + \left[ \frac{0.063137\,\i}{(\dOz+1-\i)^3} -\frac{0.012845+0.012990\,\i}{(\dOz+1-\i)^2}
+\frac{0.044414}{\dOz+1-\i} \right]\muK^2 \nn
& + \left[ \frac{0.015864}{(\dOz+1-\i)^4} +\frac{-0.0065282+0.0064556\,\i}{(\dOz+1-\i)^3} \right. \nn & \left. +\frac{0.0034840-0.022452\,\i}{(\dOz+1-\i)^2} \right. \nn & \left. +\frac{-0.0066618+0.0033960\,\i}{\dOz+1-\i} -0.0022492 \right]\frac{\muK^3}{\dOz} \nn  
& + {\cal O}[\muK^4] ~. 
\end{align}
We only wrote down the beginning of our results and with only five significant digits, 
since the space needed for further orders inflates rapidly.
But the method that we used to compute the Ernst potential allows us
to generate the series beyond ten orders with more then ten significant digits
in a reasonable amount of time; e.g.\ a personal computer with a 2.2 GHz CPU takes around 2 minutes
to compute all series up to ten orders and ten significant digits.
In Fig.~\ref{f_series_zeta_tilde}, the series of the Ernst potential is shown
as Taylor polynomials for $n=$ 3, 6 and 9 at different positions on the axis,
and it is compared to the exact potential.
One can see that for large $\tilde{\zeta}$-distances from the disc (in Fig.~\ref{f_series_zeta_tilde}, $\dOz=$ 10 or 2),
the series seems to converge for any value of $\mu$.
Closer to the disc ($\dOz=0.5$ in Fig.~\ref{f_series_zeta_tilde}),
the series does not converge any more for values of $\mu$ too much smaller than $\mu_0$.
One can also see that for very relativistic discs ($\mu>2$), the series with orders like $n=$ 3 or 6 give excellent approximations as long as the series is evaluated for sufficiently large $\dOz$.

The Taylor series using 
the other normalized and dimensionless coordinate, $\zsM\equiv \zeta/M$, 
reads as follows:
\begin{align}\label{series-of-f-disk-hat}
f(\mu;\zsM) =~& \frac{\zsM-1-\i}{\zsM+1-\i} - \frac{0.025836\,\i}{(\zsM+1-\i)^2}\muK^2 \\
&+ \frac{-0.0062737 +0.0032261\,\i\,\zsM}{\zsM\,(\zsM+1-\i)^2}\muK^3 + {\cal O}[\muK^4] ~.\nonumber
\end{align}

Another useful representation is related to Eqs~\eqref{decomposition-of-f} -- taking into consideration  that $\e^{V_0}={\cal O}[\muK]$:
\begin{equation}\label{ErnstA}
f(\mu;\zsM)   =  f_{\rm Kerr}(\mu;\zsM) \;+\; R(\mu;\zsM)
\end{equation}
with
\begin{align*}
f_{\rm Kerr}(\mu;\zsM) =~& \frac{\zsM-1-\i \hat{J}(\mu)}{\zsM+1-\i \hat{J}(\mu)} \, , \quad
\hat{J}(\mu)=\frac{-b_0(\mu)-2\Orz(\mu)\, c_1(\mu)}{\left[b_0(\mu)+ \Orz(\mu)\, c_1(\mu)\right]^2} ~,\\
R(\mu;\zsM) =~& \frac{-0.0062737}{\zsM\,(\zsM+1-\i)^2}\,\muK^3 + {\cal O}[\muK^4] ~.
\end{align*}
Note that
\begin{equation*}
\hat{J}(\mu)=1 + 0.012918\, \muK^2 - 0.0016130\, \muK^3 + {\cal O}[\muK^4] ~.
\end{equation*}

The three representations of the series of the disc's Ernst potential given above show that the solution near the limit differs from the extreme Kerr solution with the mass
$M_{\rm BH}=1/2\Omega$ by a term of order $\muK$ [see \eqref{series-of-f-disk-tilde}], from the extreme Kerr solution with the same mass $M$ by a term of order $\muK^2$ [see \eqref{series-of-f-disk-hat}] and from the Kerr solution with the same mass $M$ and angular momentum $J$ by a term of order $\muK^3$ [see \eqref{ErnstA}]. 

\section{Discussion}
The expansion of the axis potential $f(\mu;\zsM)$ of the disc solution as given in \eqref{series-of-f-disk-hat} has a remarkably simple structure: the contributions at each order of $\muK$ are given by rational functions of $\hat{\zeta}$. However, the most interesting aspect of our results is the fact that the solution near the limit is approximated very well by the (hyperextreme) Kerr solution with the same $M$ and $J$ as the disc -- up to a residual term of order ${\cal O}(\e^{3V_0})$, see \eqref{decomposition-of-f} and \eqref{ErnstA}. A precise formulation of this statement is that all multipole moments differ from those of the Kerr solution with the same $M$ and $J$ only by terms of order ${\cal O}(\e^{3V_0})$. The fact that the Kerr black hole, which is the unique stationary black hole surrounded by an asymtotically flat vacuum region, is completely characterized by the two parameters $M$ and $J$ is called the ``no-hair theorem''. Accordingly, all multipole moments of the Kerr spacetime are unique functions of $M$ and $J$. When approaching the black hole limit of the disc, the deviations of the multipole moments from those of the corresponding Kerr spacetime (same $M$ and $J$) decay $\propto \e^{3V_0}$. Thus one can say, in a sense, that the ``hair'' vanishes in this way. As already stressed in subsection \ref{hair} this seems to hold for all fluid bodies in equilibrium that permit a black hole limit. We also mention that, according to our experience, the exterior spacetime of a uniformly rotating fluid body is never described by the Kerr metric -- except for the black hole limit ($\e^{V_0}\to 0$). As long as 
$\e^{V_0}\ne 0$, all ``higher'' multipole moments (beyond $M$ and $J$) always seem to be greater than those of the corresponding Kerr spacetime \cite{fk}. 

\begin{acknowledgements}
The authors wish to thank David Petroff for many valuable discussions. This research was supported by the Deutsche Forschungsgemeinschaft (DFG) through the SFB/TR7 ``Gravitations\-wellenastronomie''.
\end{acknowledgements}

\end{document}